\documentclass[aps,pre,twocolumn,groupedaddress,showpacs]{revtex4}

\usepackage[dvips]{graphicx}
\usepackage[ansinew]{inputenc}
\usepackage{amsmath} 
\usepackage{amssymb}
\usepackage{latexsym}
\usepackage{lscape}

\newcommand{\ud}{\mathrm{d}}

\begin{document}

\newtheorem{definition}{Definition}

\title{An extended formalism for preferential attachment\\ in heterogeneous complex networks}

\author{A. Santiago}
\email[]{antonio.santiago@upm.es}
\author{R. M. Benito}
\email[]{rosamaria.benito@upm.es}
\affiliation{Grupo de Sistemas Complejos,
   Departamento de F\'\i sica y Mec\'anica,
   Escuela T\'ecnica Superior de Ingenieros Agr\'onomos,
   Universidad Polit\'ecnica de Madrid,
   28040 Madrid, Spain.}
\date{Received \today}

\begin{abstract}
In this paper we present a framework for the extension of the preferential attachment (PA) model to heterogeneous complex networks. We define a class of heterogeneous PA models, where node properties are described by fixed states in an arbitrary metric space, and introduce an affinity function that biases the attachment probabilities of links. We perform an analytical study of the stationary degree distributions in heterogeneous PA networks. We show that their degree densities exhibit a richer scaling behavior than their homogeneous counterparts, and that the power law scaling in the degree distribution is robust in presence of heterogeneity.
\end{abstract}

\maketitle



A complex network is a set of nodes and links with a non-trivial topology \cite{str01}. In the current effort to achieve a single coherent framework for complex systems, network theory has focused on the underlying principles that govern their topology \cite{alb02, new03a}. Dynamical network models \cite{dor02a} are stochastic discrete-time dynamical systems that evolve networks by the iterated addition/subtraction of nodes/links. These models regard network topology as an emergent property of the network evolution, focusing on the mechanisms that take place on such process. Among these mechanisms, preferential attachment (PA) \cite{pri65} enjoys a foremost position in network literature. 


The PA model by Barabási and Albert \cite{bar99a} has provided a minimal account of mechanisms for the emergence of \emph{scale-free} networks \cite{bar99b, dor00b}. Such networks are characterized by a degree distribution according to a power law, $P(k) = k^{-\gamma}$, which leads to a non-negligible presence of \emph{hubs} (highly connected nodes). The PA model assumes two mechanisms: growth and preferential attachment. The process starts with a seed and a new node is added to the network at each step. Each new node has a number $m$ of links attached, which are connected to the existing nodes following the so-called \emph{attachment rule}: the linking probability of a network node $v_i$ is proportional to its degree $k_i$, $\Pi (v_i) = k_i / \sum_j k_j$. This step is iterated until a number $N$ of nodes have been added. 

The PA model is strictly topological as the node degrees $\{k_i\}$ are the only metrics that drive network evolution. Nevertheless, the assumptions implicit in the PA model are not valid for a wide class of complex systems. Often, interactions between individual elements are mediated by their intrinsic properties. Although network theory has led to a significant improvement in our understanding of complex systems, it has been argued that its framework should be augmented in order to improve our modelling of complexity \cite{str01, ama04}. We will refer to \emph{heterogeneous networks} as networks where node intrinsic properties induce affinities in their interactions. We consider such networks as a logical first step in addressing the complication introduced by the influence of individual elements on the network structure. 

In recent years several dynamical models have incorporated the influence of element properties. These include weighted networks \cite{yoo01, barr04, bart04}, social models \cite{bog04, gra06}, competition models \cite{bia01, erg02}, a local knowledge model \cite{gom04}, a metric model \cite{soa05}, a gas-like model \cite{thu05} and network automata \cite{alo06}. In this letter we propose a general class of heterogeneous PA models where a function measures the \emph{affinity} between node intrinsic properties (\emph{states}) and biases the attachment probabilities of links in a generalized PA rule. The introduction of affinities aims to provide a more realistic modelling of the structure of numerous real systems that exhibit biased interactions. The proposed class contains previous models as particular cases, and provides a framework for the systematic analysis of the influence of heterogeneity in PA networks.


The PA model can be easily generalized to heterogeneous networks by imposing a metric structure on the node states, while preserving the basic mechanisms of growth and preferential attachment of the original Barabási-Albert model. Next we formally define a general class of heterogeneous PA models:
\begin{definition} A heterogeneous PA model with global affinity $M_1$ is a $3$-tuple $(R,\rho,\sigma)$, where: 
(1) $R$ is an arbitrary metric space. The elements $x \in R$ are the node \textbf{states}.
(2) $\rho : R \mapsto \mathbb{R}^+ \cup \{0\}$ is a nonnegative function with unit measure over $R$ referred as \textbf{node state distribution}.
(3) $\sigma : R^2 \mapsto \mathbb{R}^+ \cup \{0\}$ is a nonnegative function referred as \textbf{affinity} of the interactions.
\end{definition}
The class of heterogeneous models with global affinity ($HPA_g$) is the set of all $3$-tuples that satisfy the conditions in Definition 1. This formalism defines the evolution of a network according to the following rules:

\noindent (i) The nodes $v_i$ are characterized by their state $x_i \in R$. The node states describe intrinsic properties deemed constant in the timescale of evolution of the network.

\noindent (ii) The growth process starts with a seed composed by $N_0$ nodes (with arbitrary states $x_i \in R$) and $L_0$ links.

\noindent (iii) A new node $v_a$ (with $m$ links attached) is added to the network at each iteration. The number $m$ is common for all the added nodes and remains constant during the evolution of the network. The newly added node is randomly assigned a state $x_a$ following the distribution $\rho(x)$.

\noindent (iv) The $m$ links attached to $v_a$ are randomly connected to the network nodes following a distribution $\{\Pi(v_i)\}$ given by an extended \emph{attachment rule},
\begin{equation}
\label{eq:1}
\Pi (v_i) = \frac{\pi(v_i)}{\sum_j \pi(v_j)}, \qquad \pi (v_i) = k_i \cdot \sigma(x_i, x_a).
\end{equation}
The visibility $\pi$ of a node $v_i$ in the attachment rule is given by the product of its degree $k_i$ and its affinity $\sigma$ with the newly added node $v_a$, which is itself a function of the states $x_i$ and $x_a$. It thus can be seen that for each interaction $\sigma$ biases the degree $k_i$ of the candidate node. Steps (iii) and (iv) are iterated until a desired number of nodes has been added to the network. To sum up, the choice of the tuple $(R, \rho, \sigma)$ determines the form of heterogeneity in the attachment mechanism.


Next we derive an analytical solution for the stationary degree distribution $P(k)$ of the class $HPA_g$ of heterogeneous models. The solution is obtained by \emph{rate equations} which establish a balance in the flows of degree densities over a partition of a growing network. Let us first define a sequence of functions $\{f(k,x,N)\}_{N>0}$ which measure the probability density of a randomly chosen node having degree $k \in \mathbb{N}$ and state $x \in R$ in a network at the iteration $t = N$. The degree densities are local metrics, thus they uniformly converge when $N \to \infty$ to a stationary density function $f(k,x)$. Finally, the stationary degree distribution $P(k)$ measures the probability of a randomly chosen node having degree $k$ in the thermodynamic limit.

The rationale for the choice of $f$ in modelling the evolution of the degrees is simple: in a heterogeneous model, nodes with equal degree may contribute differently to changes in $P(k,N)$ according to their state, despite being homogeneously accounted by $P$. Thus, for each $(k,N)$ we need to know the influence of $x$ on the degree distribution. We will denote by $V(x) = \{v_i , x_i = x\}$ the subset of nodes in the network with state $x$. Assuming that the assignation of states $x_i$ is uncorrelated with the topology of the growing network, and that there are no linking events between existing nodes, the sequence $\{f\}$ can be modelled on each $V(x)$. For each $x$, the form of the equation will be $L_1 - L_2 = R_1 - R_2$, where:\\
$L_1 =$ density of nodes with degree $k$ at $t = N+1$;\\
$L_2 =$ density of nodes with degree $k$ at $t = N$;\\
$R_1 =$ increase in density due to nodes with degree $k-1$ that have gained a link at $t = N$;\\
$R_2 =$ decrease in density due to nodes with degree $k$ that have gained a link at $t = N$.

The total density of nodes with state $x$ and degree $k$ at $t = N+1$ is $(N+1)f(k,x,N+1)$. Here we assume that $N$ is large and no incoming node is rejected due to lack of affinity, therefore we approximate the network size by $t$. Likewise, the total density of nodes with state $x$ and degree $k$ at $t = N$ is $Nf(k,x,N)$. In order to estimate the changes in $f$ due to the adquisition of new links, we introduce a mean-field \emph{fitness} factor $w(x)$ that measures the average affinity of a network node with state $x$ towards the incoming nodes, 
\begin{equation}
w(x) \equiv \int_R \sigma(x,y) \rho(y) \ud y.
\end{equation}
The probability of a node being chosen for attachment is proportional to $k \, w(x)$, and the decrease in the density of nodes with state $x$ and degree $k$ due to their attachment of a new link is proportional to $k \, f(k,x,N) \, w(x)$. This term must be normalized by the increase in all the densities (irrespective of their degree $k$ and state $x$) in the network at $t = N$, measured by a partition factor
\begin{equation}
\psi(N) \equiv \sum_k k \int_R f(k,x,N) w(x) \ud x.
\end{equation}
As in each iteration there are $m$ links attached to a newly added node, the decrease in the total density of nodes with state $x$ and degree $k$ due to nodes with degree $k$ gaining a link and being promoted to degree $k+1$ is $m \, k f(k,x,N) w(x) / \psi(N)$ at each step. Here we assume that for large $N$ the probability that a single node receives more than one link attached to the same new node becomes negligible. Likewise, the increase in the total density of nodes with state $x$ and degree $k$ due to nodes with state $x$ and degree $k - 1$ gaining a link and being promoted to degree $k$ is $m(k-1) f(k-1,x,N) w(x) / \psi(N)$. The resulting density rate equation for $k > m$ is:
\begin{eqnarray}
(N+1)f(k,x,N+1) - Nf(k,x,N) =\nonumber\\
= \frac{mw(x)}{\psi(N)}[(k-1)f(k-1,x,N) - k f(k,x,N)].
\end{eqnarray}

When $k = m$, there is no increase in $f(m,x,N)$ due to the promotion of nodes with $k = m-1$. However, each newly added node has degree $m$ and this contribution to $f(m,x,N)$ is distributed according to $\rho(x)$, thus:
\begin{eqnarray}
(N+1)f(m,x,N+1) - Nf(m,x,N) =\nonumber\\
= \rho(x) - \frac{m \, w(x)}{\psi(N)} m \, f(m,x,N).
\end{eqnarray}
There are no nodes with degree $k < m$, since when $N \to \infty$ all the links attached to newly added nodes find receptive nodes in the network, therefore the previous equations define all the possible cases in each iteration.

In the thermodynamic limit $N\to \infty$, $f(k,x,N+1) = f(k,x,N) = f(k,x)$ and the rate equations become:
\begin{eqnarray}
\label{eq:coupledsys}
& f(k,x) = \\
& \left\{ \begin{array}{ll}
\frac{mw(x)}{\psi} [(k-1)f(k-1,x) - kf(k,x)] & \textrm{for } k > m,\\
\rho(x) - \frac{m w(x)}{\psi} m f(k,x) & \textrm{for } k=m,
\end{array} \right. \nonumber
\end{eqnarray}

These equations are coupled by the stationary partition factor $\psi \equiv \sum_k k \int f(k,x) \, w(x) \, \ud x$. Notice that in the homogeneous PA model, $w(x) = 1 \, \forall x \in R$ and thus $\psi = \sum_k k P(k)$, which is equal to the average network degree $z = (2Nm + 2L_0)/(N + N_0)$. In the stationary limit $N \gg N_0$ and $Nm \gg L_0$, thus $\psi \simeq 2m$.

To the contrary, in the heterogeneous models $w(x)$ continuously weighs the density $f(k,x)$ in the integral associated to each degree $k$ in the stationary partition factor $\psi$. In order to decouple Eq. \ref{eq:coupledsys}, let us assume that $w(x)$ doesn't change too much on $R$ so that we may approximate $w(x)$ by its mean $\bar{w} \equiv \int_R w(x) \rho(x) \ud x$, then
\begin{equation}
\psi \simeq \sum_k k \int f(k,x) \, \bar{w} \, \ud x = \sum_k k \, P(k) \bar{w} \simeq 2m \, \bar{w}. 
\end{equation}

The resulting decoupled system becomes:
\begin{eqnarray}
\label{eq:finalrates}
& f(k,x) =\\
& \left\{ \begin{array}{ll}
\frac{1}{2} \left( \frac{w(x)}{\bar{w}} \right) [(k-1)f(k-1,x) - kf(k,x)] & \textrm{for } k > m, \nonumber\\
\rho(x) - \frac{1}{2} \left( \frac{w(x)}{\bar{w}} \right) m f(k,x) & \textrm{for } k=m,
\end{array} \right.
\end{eqnarray}

We introduce a \emph{normalized fitness} $\hat{w} = w(x)/\bar{w}$, which measures the fitness of a state relative to the average fitness of the network nodes. Solving Eq. \ref{eq:finalrates} we obtain
\begin{eqnarray}
\label{eq:solutions}
f(m,x) & = & \frac{2\rho}{\hat{w}m+2}, \\
f(k,x) & = & \frac{\hat{w}(k-1)}{\hat{w}k+2} f(k-1,x) \;\, \textrm{for } k > m. \nonumber
\end{eqnarray}
Thus the solution for the stationary density for $k>m$ is
\begin{equation}
\label{eq:density}
f(k,x) = \left( \prod_{j = m+1}^k \frac{\hat{w}(j-1)}{\hat{w}j+2} \right) \frac{2\rho}{\hat{w}m+2}
\end{equation}
and integrating the density in Eq. \ref{eq:density} over the state space $R$ the stationary degree distribution is
\begin{equation}
\label{eq:distrib}
P(k) = \int_R \left( \prod_{j = m+1}^k \frac{\hat{w}(j-1)}{\hat{w}j+2} \right) \frac{2\rho}{\hat{w}m+2} \ud x.
\end{equation}
It is important to remark that the solutions in Eqs. \ref{eq:density} and \ref{eq:distrib} are valid for any model in our class, since they do not make any assumption regarding the geometry of space $R$, the form of affinity $\sigma$ or the distribution $\rho$.

Furthermore we may assume that $w(x) > 0 \; \forall x \in R$ without loss of generality, otherwise $\sigma(x_0,y) = 0 \; \forall y \in R$ since $\sigma \geq 0$, which would mean that $x_0$ is ``inert'' and may be excluded from $R$. Then $\hat{w}(x) > 0 \; \forall x \in R$ and we may write the density in Eq. \ref{eq:density} for $k>m$ as
\begin{equation}
\label{eq:densitybeta}
f(k,x) = \frac{2\rho/\hat{w}}{m+2/\hat{w}} \frac{B(k, 1+2/\hat{w})}{B(m, 1+2/\hat{w})}.
\end{equation}
where Legendre's Beta function $B(y,z) = \int_0^1 t^{y-1} (1-t)^{z-1} \ud t$ for $y,z>0$ satisfies the functional relation $\Gamma(a) / \Gamma(a+b) = B(a,b) / \Gamma(b)$ for Euler's Gamma $\Gamma$ function. Likewise, integrating the density in Eq. \ref{eq:densitybeta} we obtain for $k>m$ the stationary degree distribution
\begin{equation}
\label{eq:distrib2}
P(k) = \int_R \frac{2\rho/\hat{w}}{m+2/\hat{w}} \frac{B(k, 1+2/\hat{w})}{B(m, 1+2/\hat{w})} \ud x. 
\end{equation}

\begin{figure}
\centering \includegraphics[width=0.48\textwidth]{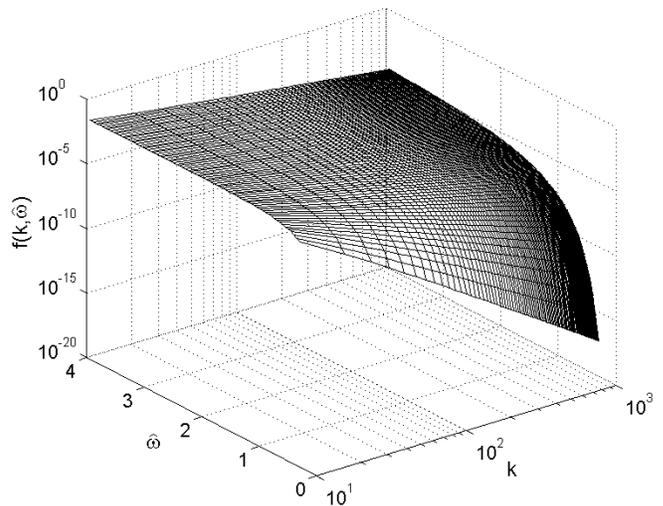}
\caption{Solution for the stationary density $f(k,\hat{w})$ in heterogeneous PA models (Eq. \ref{eq:density}) with $\rho(x) = 1$ and $m = 3$.}
\label{fig:1}
\end{figure}

Notice that Eq. \ref{eq:densitybeta} reduces to $f(m,x)$ in Eq. \ref{eq:solutions} when $k=m$, thus Eqs. \ref{eq:densitybeta} and \ref{eq:distrib2} are valid for $k \ge m$. Also, notice that the stationary densities $f(k,x)$ in the heterogeneous formalism, as given by Eq. \ref{eq:densitybeta}, follow the form of a Beta function with arguments $(k, 1+2/\hat{w})$.

In particular cases it is possible to obtain simpler expressions for $f(k,x)$. First, in the homogeneous case, $w(x) \equiv 1$, thus $\hat{w} = w = 1$ and Eq. \ref{eq:density} yields
\begin{equation}
f(k,x) = \frac{2\rho \, m(m+1)}{k(k+1)(k+2)},
\end{equation}
and the density functions $f(k,x)$ asymptotically follow power laws $f(k,x) \sim k^{-\gamma}$ with exponent $\gamma = 3$ irrespective of the node states. This is reasonable, as the dynamics of the homogeneous PA network is not affected by them, thus the asymptotic behavior of all density components is the same. If furthermore we assume that $\rho(x)$ is uniform over $R$, then any subset $V(x)$ of nodes will exhibit the same distribution as the whole network. Likewise, the degree distribution $P(k)$ asymptotically follows a power law with exponent $\gamma = 3$ like the Barabási-Albert model \cite{dor00b}, irrespective of the distribution $\rho$.

In presence of heterogeneity the fitness $w(x)$ is generally not constant over $R$. It is still possible to obtain simple forms of $f(k,x)$ from Eq. \ref{eq:density} for discrete values of $\hat{w}$ that render the denominator a product of integers. For instance, let $x$ be such that $\hat{w} = 2 > 1$, then
\begin{equation}
f(k,x) = \frac{\rho \, m}{k(k+1)}, 
\end{equation} 
which asymptotically behaves as a power law with exponent $\gamma = 2$. Conversely, letting $x$ be such that $\hat{w} = 2/3 < 1$, then Eq. \ref{eq:density} yields
\begin{equation}
f(k,x) = \frac{3\rho \, m(m+1)(m+2)}{k(k+1)(k+2)(k+3)}, 
\end{equation}
which asymptotically behaves as a power law with exponent $\gamma = 4$. By extension, normalized fitness values $\hat{w} = 2/n$ with integer $n$ yield density functions $f(k,x)$ with an asymptotic power law behavior with $\gamma = 1 + 2/\hat{w}$.

More generally, it can be proved that the Beta function behaves as $B(y,z) \sim y^{-z}$ when $y \to \infty$. Therefore, Eq. \ref{eq:densitybeta} implies that $f(k,x)$ behaves asymptotically as $B(k,1 + 2/\hat{w}) \sim k^{-(1 + 2/\hat{w})}$. The density components $f(k,x)$ in heterogeneous PA networks exhibit a \emph{multiscaling} according to power laws $k^{-\gamma(\hat{w})}$ along the continuous spectrum of normalized fitness $\hat{w}$. The scaling exponents are inversely related to $\hat{w}$ according to $\gamma(\hat{w}) = 1 + 2/\hat{w}$ so that they span themselves a continuum. 

Thus, as $\hat{w}$ increases (resp. decreases), the exponent $\gamma$ of $f$ decreases (resp. increases). Nodes with states more fit than the average ($\hat{w} > 1$) adopt densities with $\gamma < 3$, which exhibit a slower asymptotical decay, and tend to produce more hubs. Nodes with states less fit than the average ($\hat{w} < 1$) adopt densities with $\gamma > 3$, which exhibit a faster asymptotical decay. These results are verified in Fig. \ref{fig:1}, which shows the plot of the density $f$. Although the shape of $f$ depends on the particular heterogeneity of the network, its asymptotic behavior can be characterized in a general way by plotting the dependence of $f$ on $\hat{w}(x)$ instead of $x$.

The behavior described points out to a \emph{signature} of heterogeneity in the topology of PA networks: the fitness $w(x)$ will tend to fluctuate on $R$ and the density components $f(k,x)$ will exhibit different scaling exponents. This prompts us to suggest a procedure for empirically detecting such signature in real networks. In order to check the role of a certain degree of freedom in the network evolution, one could select different states $\{x\}$ varying only in such dimension, select subsets of nodes $V_{\delta}(x) = \{v_i , d(x^{(i)} - x) < \delta\}$ with states in a small neighborhood of the states $x$. If the partial distributions $\{P(k | x)\}$ exhibit power laws with different exponents, it would be feasible to assume this dimension (or another one correlated with it) is biasing the attachment mechanism. To check the existence of heterogeneity it would be enough to find one such subset of nodes whose partial distribution $\{P(k | x)\}$ exhibits a power-law with an exponent different to the one in $P(k)$.

As for the degree distribution $P(k)$ of the heterogeneous models, Eq. \ref{eq:distrib}, it is obtained by the integration of the density components which form a set of power-laws with varying exponents. The level of relative variability present in the fitness $w(x)$ over the state space $R$ will determine the ultimate composition of $P(k)$. Choices of $(R,\rho,\sigma)$ that yield little fluctuations in $w(x)$ on $R$ will translate into density functions $f(k,x)$ deviating slightly from the homogeneous power law, and degree distributions similar to the Barabási-Albert case. On the other hand, choices of $(R,\rho,\sigma)$ that yield larger fluctuations in $w(x)$ will translate into a wider spectrum of exponents $\gamma$ in the density components, and thus a larger deviation of the degree distribution from the homogeneous case. 

While we need to know the tuple $(R,\rho,\sigma)$ to obtain the distribution $P(k)$, it is still possible to make rough predictions about its behavior in heterogeneous PA networks. We may regard the degree distribution in such networks as a superposition of power laws whose exponents are distributed around the homogeneous $\gamma = 3$. For low degrees it can be expected that averaging low and high $\gamma$ components will tend to produce a behavior of $P(k)$ similar to the one observed in homogeneous networks. When $k \to \infty$, the tail of $P(k)$ will be dominated by the slowest decaying components, associated to states with highest $w$. Furthermore, given that $\hat{w}$ will be distributed by definition around $1$, the $P(k)$ of all the models in the class will exhibit scaling exponents satisfying $1 < \gamma \le 3$. This is remarkable since the empirical observations in real scale-free networks show that most exponents fit within this interval.


In summary, we have proposed a generalization of the PA model to heterogeneous networks. We have defined a general class of heterogeneous PA models where node properties are described by fixed states in an arbitrary space, and where an affinity function biases the attachment probabilities of links. We have obtained analytical solutions for the degree densities and degree distribution of heterogeneous networks in the stationary limit. We have shown that these networks exhibit a richer scaling behavior than their homogeneous counterparts, a multiscaling of their degree densities according to power laws with exponents spanning a continuum. Such phenomenon leaves a signature of heterogeneity in the topology of PA networks that can be empirically checked in real networks. We have also shown the relationship between the variability in the exponents and the fluctuations in the fitness over the state space, and the consequences of this on the asymptotic behavior of the degree distribution. Finally we have shown that the degree distributions of all heterogeneous models in our class exhibit power laws with exponents within the limits empirically observed in real networks. These results suggest that the ubiquity of real scale-free networks with otherwise different structural properties may be partially explained by PA mechanisms with varying levels of heterogeneity.

This work has been supported by the Spanish MEC under Project 'i-MATH' No.~CSD2006-00032 and Project No.~MTM2006-15533, and GESAN, S.A.


\end{document}